\documentclass[prl,twocolumn,showpacs,floatfix,superscriptaddress,amsfonts]{revtex4}
\usepackage{graphicx,graphics,color,epsfig}% Include figure files
\usepackage{bm}
\usepackage{amsmath}
\usepackage{amssymb}
\usepackage{epstopdf}
\usepackage{color}

\begin{document}
\preprint{}
\title{Band Narrowing and Mott Localization in Iron Oxychalcogenides  La$_2$O$_2$Fe$_2$O(Se,S)$_2$}
\author{Jian-Xin Zhu}
\affiliation{Theoretical Division, Los Alamos National Laboratory,
Los Alamos, New Mexico 87545, USA}
\author{Rong Yu}
\affiliation{Department of Physics \& Astronomy, Rice University, Houston, Texas 77005, USA}
\author{Hangdong Wang}
\affiliation{Department of Physics, Zhejiang University, Hangzhou 310027, P. R. China}
\author{Liang L. Zhao}
\affiliation{Department of Physics \& Astronomy, Rice University, Houston, Texas 77005, USA}
\author{M. D. Jones}
\affiliation{University at Buffalo, SUNY, Buffalo, New York 14260, USA}
\author{Jianhui Dai}
\affiliation{Department of Physics, Zhejiang University, Hangzhou 310027, P. R. China}
%\affiliation{Zhejiang Institute of Modern Physics, Zhejiang University, Hangzhou 310027, P. R. China}
\author{Elihu Abrahams}
\affiliation{Center for Materials Theory, Rutgers University, Piscataway, New Jersey 08855, USA}
\altaffiliation{Current address: Department of Physics and Astronomy, University of California Los Angeles, Los Angeles, California 90095, USA.}
\author{E. Morosan}
\affiliation{Department of Physics \& Astronomy, Rice University, Houston, Texas 77005, USA}
\author{Minghu Fang}
\affiliation{Department of Physics, Zhejiang University, Hangzhou 310027, P. R. China}
%\affiliation{Zhejiang Institute of Modern Physics, Zhejiang University, Hangzhou 310027, P. R. China}
\author{Qimiao Si}
\affiliation{Department of Physics \& Astronomy, Rice University, Houston, Texas 77005, USA}

\date{\today}
\begin{abstract}
Bad metal properties have motivated a description of the parent iron
pnictides as correlated metals on the verge of Mott localization.
What has been unclear is whether interactions can push
these and related compounds to the Mott insulating side of the
phase diagram. Here we consider the iron oxychalcogenides
La$_2$O$_2$Fe$_2$O(Se,S)$_2$, which contain an Fe square lattice
with an 
%\qsx{enlarged}
expanded
unit cell. We show theoretically that they
contain enhanced correlation effects through 
band narrowing
%\qsx{band narrowing}
%\qs{narrowing of Fe 3d-electron bands}
compared to LaOFeAs, and we provide experimental evidence that
they are Mott insulators with moderate charge gaps.
We also discuss the magnetic properties in terms
of a Heisenberg model with frustrating $J_1$-$J_2$-$J_2^{\prime}$
exchange interactions on a ``doubled'' checkerboard lattice.
\end{abstract}
\maketitle

{\it Introduction:~}
Iron pnictides are the first non-copper-based compounds to exhibit high-$T_c$
superconductivity~\cite{Kamihara_FeAs,Zhao_CPL08}, and have therefore attracted
considerable interest.
%~\cite{Day_physicstoday_aug2009}.
How strong the electron correlations are in the parent
iron pnictides is a subject of extensive current discussion.
One viable description places these compounds at the boundary of
itinerancy and Mott localization.
The motivation for this incipient Mott picture
comes from the observed bad metal properties~\cite{Si-Abrahams-prl08,Si-Abrahams-Dai-Zhu},
and from first-principles calculations~\cite{Haule-Kotliar-prl08,Laad08}
and related~\cite{Daghofer08,CFang08,CXu08} considerations.
Optical conductivity has in the meantime shown
a sizable suppression of the Drude weight \cite{Qazilbash,Hu08,Si_natphys},
as well as temperature-induced spectral-weight transfer going to energies
of the eV range \cite{Hu08,Yang08,Boris09}. Inelastic neutron scattering
experiments have provided complementary support.
They have not only observed zone-boundary high-energy spin waves \cite{Zhao},
%\onlinecite{Zhao}.
%They have in addition
but also shown~\cite{Zhao,Diallo}
that the total
(ordered plus fluctuating)
spin spectral weight is, for instance, about $1.2$ $\mu_B$/Fe
in CaFe$_2$As$_2$.
%\qsx{The latter}
Such a large spin spectral weight
 implies that the low-energy spin degrees of freedom come
from electronic excitations not only close to the Fermi energy
but also far away from it.
%far away from the Fermi energies. 
All these are
defining properties of metals on the
%boundary between itinerancy and Mott localization.
verge of Mott localization. To establish the incipient Mott picture,
however, it is important to identify the Mott insulating part of the
electronic phase diagram in the iron pnictides and related compounds.

\begin{figure}[t]
\centerline{\psfig{figure=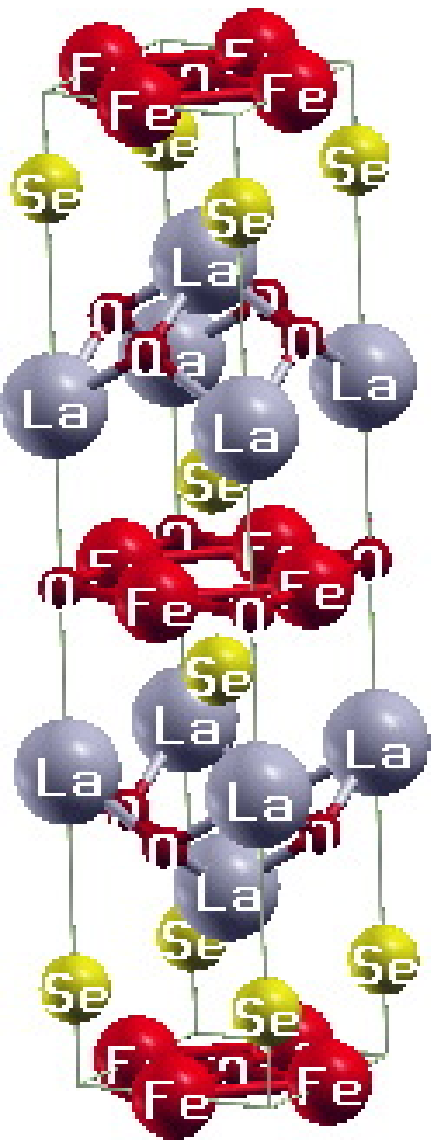,width=2.1cm,angle=0}\psfig{figure=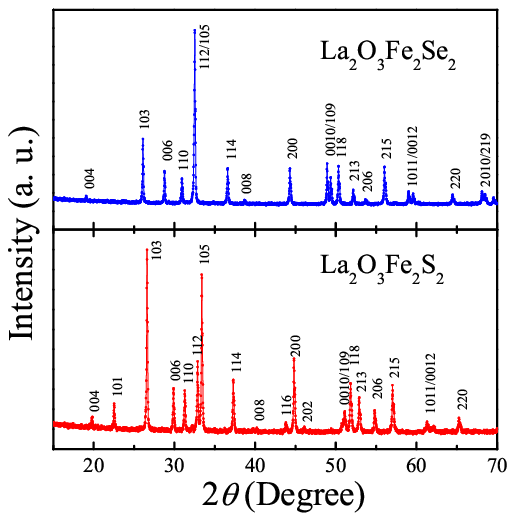,width=5.9cm}}
\caption{(Color online) (Left panel) Crystal structure of
La$_2$O$_2$Fe$_2$OSe$_2$; (Right panel) XRD patterns
for the La$_2$O$_3$Fe$_2$Se$_2$ and La$_2$O$_3$Fe$_2$S$_2$
samples. 
The cell parameter values are $a=b=4.085$ \AA\; and $c=18.605$ \AA\;  
for La$_2$O$_3$Fe$_2$Se$_2$ and $a=b=4.042$ \AA\;  and $c = 17.929$ \AA\; 
for La$_2$O$_3$Fe$_2$S$_2$. 
Atomic positions  in the unit cell of a paramagnetic state are 
as follows: La (1/2,1/2,0.18445), Fe (0.5,0,0), Se (0,0,0.09669),
O1 (1/2,0,1/4), O2 (1/2,1/2,0).
}
\label{FIG:CrystalStructure}
\end{figure}

Towards this end, we consider the iron oxychalcogenides
La$_2$O$_2$Fe$_2$OSe$_2$
and La$_2$O$_2$Fe$_2$OS$_2$.
These systems are built from
% the
stacking
%secondary
layered
units La$_2$O$_2$ and Fe$_2$O(Se,S)$_2$.
We will describe the Se case below, with the understanding that the S case
is similar unless otherwise stated.
%In the hypothetical antiferromagnetic structure, it belongs to the space group $P42/mmc$.
Fig.~\ref{FIG:CrystalStructure} (left panel) shows the crystal structure
with space group I4/mmm as given in Ref.\ \cite{Mayer92}.
Each Fe$_2$OSe$_2$ layer contains a square lattice of Fe ions, and this is the same
as in an FeAs layer of LaOFeAs or an FeTe layer of the compound FeTe
~\cite{M.Fang,MK.Wu}.
In addition, ${\rm Fe^{2+}}$ is the nominal valence as in the latter
materials. The Fe$_2$OSe$_2$ layer
is unique in that the Fe ions in each plaquette are alternatively
linked by an in-plane oxygen ion and by two selenium ions buckled
on the two sides of the Fe square lattice.
% The parameters for the
%unit cell and internal atomic positions
%are given in the caption of Fig.~\ref{FIG:CrystalStructure}.
% Of particular interest to note here
The inter-atomic distances
are $d_{\text{Fe-Fe}}=2.884$ \AA, $d_{\text{Fe-O}}=2.039$ \AA, and
$d_{\text{Fe-Se}}=2.722$ \AA.

An important observation is that Fe-square-lattice unit cell
of La$_2$O$_2$Fe$_2$OSe$_2$
is slightly larger (by about 1\%)
than that of LaOFeAs and considerably larger
than that of either
FeTe (by over 6\%) or FeSe (by over 8\%)~\cite{M.Fang,MK.Wu}.
%FeTe_Lattice_Parameter}
This raises the possibility for a narrower bandwidth ($\propto t$)
and, correspondingly, a larger strength of the normalized electron
correlation, $U/t$. We have further been motivated by the
indication of antiferromagnetic ordering (AFM)
in the early work of Mayer and co-workers~\cite{Mayer92}.

{\it Band narrowing:~}
We have determined the electronic structure using
the WIEN2k linearized augmented plane wave method~\cite{PBlaha:2001}
based on density functional theory (DFT).
A generalized gradient approximation (GGA)~\cite{JPPerdew:1996} was used to treat exchange and correlation. 

\begin{figure}[th]
\includegraphics[width=0.9\linewidth]{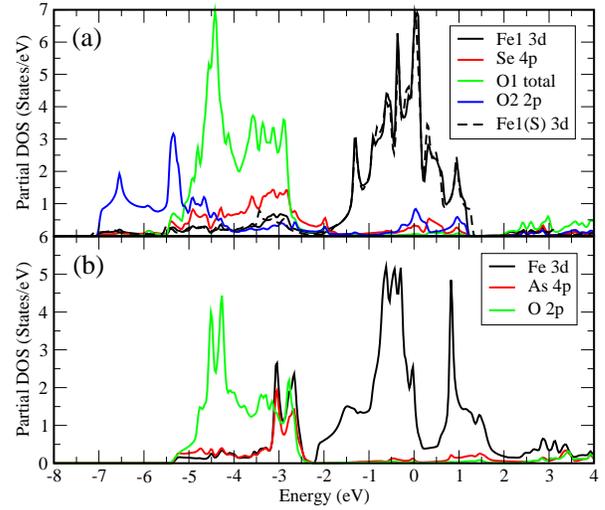}
\caption{(Color online) The partial DOS
for paramagnetic La$_2$O$_2$Fe$_2$OSe$_2$ (a) and LaOFeAs (b).
Also shown in Panel (a) is the Fe $3d$ partial DOS for
La$_2$O$_2$Fe$_2$OS$_2$.
In our DFT calculation,
the energy threshold separating
the localized and non-localized electronic
states is chosen to be $-6 \; \text{Ryd}$.
The muffin-tin radii are: 2.37 $a_0$ (Bohr radius)
for Lanthanum, 2.03 $a_0$ for iron, 2.40 $a_0$ for selenium,
and 1.80 $a_0$ for oxygen.
%,where $a_0$ is the Bohr radius.
The criterion for the number of plane waves is chosen to
be $R_{\text{MT}}^{\text{min}}*K^{\text{max}}=7$ and the
number of $k$-points is $32\times 32 \times 8$
for these paramagnetic-state calculations.
}
\label{FIG:Fe-DOS}
\end{figure}

In Fig.~\ref{FIG:Fe-DOS}, we plot the projected density of states (DOS).
In both compounds, we find that the $3d$ electrons on Fe
contribute most to the DOS near the Fermi energy.
The Fe $d$-electron DOS is mostly confined between
$-2$ eV and 1.2 eV. This represents considerable narrowing of the
Fe $d$-electron bands compared to LaFeAsO,
where it occurs
between $-2.2$ eV and $2$ eV~\cite{DJSingh:2008}.
(There are also some differences between the two systems 
in the DFT-derived DOS near the Fermi energy, but the effective
$U/t$ primarily depends on the overall 3d-electron bandwidth.)
%We have also calculated the band structure of La$_2$O$_2$Fe$_2$OS$_2$,
%finding similar results
The results for La$_2$O$_2$Fe$_2$OS$_2$ [dashed line
of Fig.~\ref{FIG:Fe-DOS}(a)] are similar.
A 
%similar 
comparable degree of 3d-electron band narrowing exists when
the iron oxychalcogenides are compared to the iron
chalcogenides FeTe and FeSe \cite{Subedi_FeTe:2008}.

{\it Mott insulating behavior:}~
The narrower Fe $3d$ electronic bands point to enhanced
correlation effects.
To explore this, we have synthesized
these materials following a similar procedure to that taken in Ref.\ \cite{Mayer92}
and measured their transport and magnetic properties.
Polycrystalline samples with nominal composition La$_2$O$_3$Fe$_2$M$_2$
(M = S or Se) were
prepared by conventional solid state reaction using high purity La$_2$O$_3$,
Fe and S (or Se) powder as starting materials.
The samples were characterized by powder
X-ray diffraction (XRD) 
(right panel of Fig.\ \ref{FIG:CrystalStructure})
with Cu K$\alpha$ radiation ($\lambda =1.5418$ \AA)
at room temperature.
%Fig.\ \ref{FIG:CrystalStructure} (right panel) shows the powder XRD patterns.

\begin{figure}[th]
\includegraphics[width=0.7\linewidth]{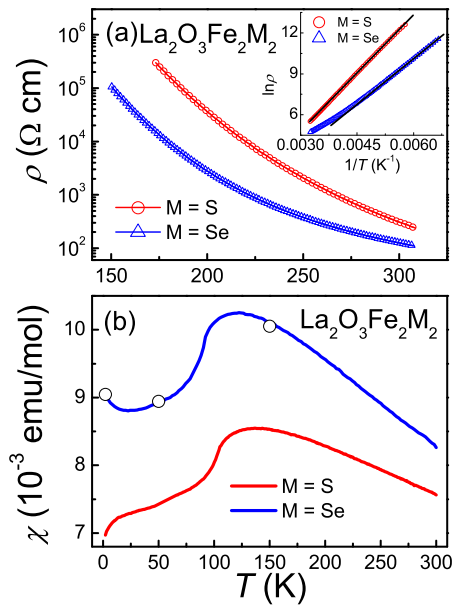}
\caption{(Color online)
(a) The electrical resistivity ($\rho$) vs. temperature ($T$) for
La$_2$O$_3$Fe$_2$S$_2$ and La$_2$O$_3$Fe$_2$Se$_2$.
Inset: ln$\rho$ vs. $1/T$;
(b) The temperature dependence of susceptibility, $\chi$, measured at 3 Tesla.
The three open points,
measured at 2 K, 50 K, and 150 K, respectively,
are determined from the slope of the M(H) curves.
}
\label{FIG:Resist-Suscep}
\end{figure}

The electrical resistivity
as a function of temperature is shown
in Fig.~\ref{FIG:Resist-Suscep}(a).
For both  M = Se and M = S,
the behavior is insulating
with activation energy gaps [{\it cf.} inset of
Fig.~\ref{FIG:Resist-Suscep}(a)]
of about 0.19 eV and 0.24 eV,
respectively.
Figure~\ref{FIG:Resist-Suscep}(b)
shows the temperature dependence of the bulk magnetic susceptibility
$\chi(T)$,
% At temperatures above 150 K, $\chi(T)$ for both samples increases 
%almost linearly with decreasing temperature and 
which
exhibits a broad maximum around 120 K. % in both cases.
The rapid decrease, at $93$ K and $105$ K, for M = Se and M = S respectively,
is naturally ascribed to antiferromagnetic ordering
in the Fe $d$-electron moments.
At the lowest measured temperatures,
$\chi(T)$ shows a  small increase for La$_2$O$_3$Fe$_2$Se$_2$
but a small decrease for La$_2$O$_3$Fe$_2$S$_2$.
This may be due to
different impurities, whose amounts must be small as
there is no trace of them in the XRD patterns.

\begin{figure}[th]
\includegraphics[width=1\linewidth]{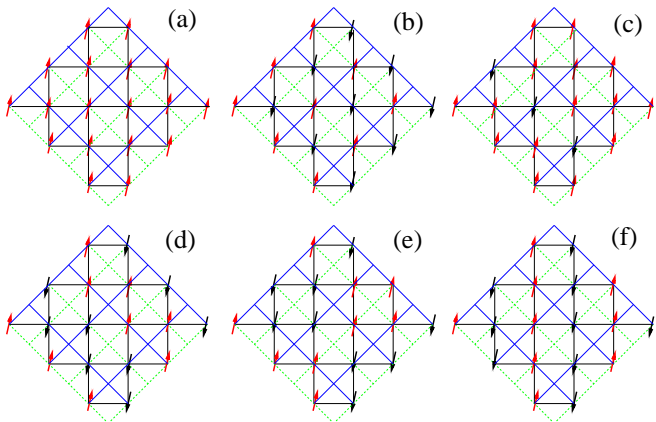}
\caption{(Color online) Schematic representation of the spin building blocks
for the magnetically ordered states of La$_2$O$_2$Fe$_2$OSe$_2$.
To be consistent with the effective model, only Fe atoms are shown.
The arrows represent the spin orientation.
The black solid, green dashed, and blue solid lines
%segments
stand for $J_1$, $J_2$, and $J'_2$ spin exchange paths.}
\label{FIG:Spin}
\end{figure}

The results provide evidence that both La$_2$O$_3$Fe$_2$Se$_2$
and La$_2$O$_3$Fe$_2$S$_2$ are Mott insulators, {\it i.e.}
correlation-induced insulators with low energy
spin excitations that order antiferromagnetically at low temperatures.

{\it Magnetism:~}
To investigate the Mott insulating state,
we extend our DFT calculations to the cases of magnetically ordered states using
the GGA+$U$ method.
We consider seven magnetic ground states,
the building blocks for which are illustrated in
Fig.~\ref{FIG:Spin}.\
%We notice from Fig.~\ref{FIG:CrystalStructure} (left panel) that each 
Each crystalline unit cell consists of two layers of Fe$_2$OSe$_2$,
%and their 
whose coordinating geometry is rotated with respect to each
other by $90^\circ$ along the $c$-axis.
Correspondingly, the seven spin states are formed with the
spin alignment as follows:
FM ((a)+(a)), AFM1 ((b)+(b)), AFM2 ((c)+(a)), AFM3 ((d)+(a)),
AFM4 ((e)+(e)), AFM5 ((f)+(f)), and AFM6 ((e)+(e)'),
where the notation ((spin1)+(spin2)) represents
only the spin configuration rather than the underlying
atomic coordination within each layer and  the symbol
(e)' represents the spin configuration obtained by
rotating the configuration (e) in Fig.~\ref{FIG:Spin}
by $90^\circ$. 
The energy differences between these
spin ordered states for various on-site Hubbard interaction
$U$ are summarized in Table~\ref{Table:Energy}.
(None of the magnetically-ordered states at $U=0$ has a gap
in the DOS at the Fermi energy. Within GGA+$U$, an insulating gap
does develop
for sufficiently large value of $U$,
for some of the magnetic ordering patterns.)
Based on considerations similar to the FeAs
case~\cite{Si-Abrahams-prl08,Yildirim,Maetal,Yinetal},
we model the spin Hamiltonian in terms of nearest-neighbor (n.n.)
exchange interaction
($J_1$) and the next-nearest-neighbor (n.n.n.)
exchange interaction across
an Fe-square plaquette with two buckled Se ions ($J_2$)
and the n.n.n. interaction across an Fe-square plaquette
containing the O-ion ($J_2'$).
Even though the spin system is expected to be largely Heisenberg-type,
the exchange parameters can be extracted by
fitting the 
%GGA+$U$ 
ground-state energies
in terms of the Ising counterpart
~\cite{DDai:2003,HKabbour:2008}.
We summarize the extracted exchange interactions
in Table~\ref{Table:Exc}. 

\begin{table}[h]
%\begin{table*}
\begin{ruledtabular}
\begin{tabular}{cccccccc}
%\hline
$U$ (eV) & FM & AFM1 & AFM2 & AFM3 & AFM4 & AFM5 &AFM6 \\
\hline
0    & 0 & 226  & 39.7   & -38.1  & -10.7 & 45.5  & -99.2 \\
1.5 & 0 & -164 & -19.4 & -69.6  & -133 & -209 & -274 \\
3.0 & 0 & -241 & -52.4 & -90.4  & -179 & -237 & -276\\
4.5 & 0 & -216 & -48.4 & -74.2 & -147 & -188& -212\\
\end{tabular}
\end{ruledtabular}
\caption{Relative energies $\Delta E$ (meV/unit cell) of the FM
(ferromagnetic) and six
%, AFM1, AFM2, AFM3, AFM4, AFM5, and AFM6 
AFM states of 
La$_2$O$_2$Fe$_2$OSe$_2$ obtained from GGA+$U$ calculations.}
\label{Table:Energy}
%\end{table*}
\end{table}

\begin{table}[h]
\begin{ruledtabular}
\begin{tabular}{ccccc}
$U$ (eV) & $J_{1}/S^2 $ & $J_{2}/S^2$ & $J'_{2}/S^2 $ & $\sigma$ \\
\hline
0    & -3.53 & 2.10  & 9.23 & 139.5 \\
1.5 & 2.53 & -2.40 & 6.17  & 5.5 \\
3.0 & 3.77 & -1.11 & 4.79  & 1.27 \\
4.5 & 3.38 & -0.78 & 3.28 & 0.63
\end{tabular}
\end{ruledtabular}
\caption{Values of spin exchange parameters $J_1/S^2$, $J_2/S^2$, $J'_2/S^2$
(meV) for La$_2$O$_2$Fe$_2$OSe$_2$ estimated from GGA+$U$ calculations.
Positive (negative) $J$'s
correspond to antiferromagnetic (ferromagnetic) exchange.
The last column lists the ``error bar" in energy (meV/unit cell): 
$\sigma = [\sum_{i=1}^{N}(\Delta E^{(i)} 
- \Delta E_{M}^{(i)})^{2}/(N-1)]^{1/2}$, where $N=6$
and $\Delta E^{(i)}$ and $\Delta E_{M}^{(i)}$ are the relative 
ground-state energies in DFT and in the extracted effective spin models,
respectively. $\sigma$ is large for $U=0$, but becomes relatively
small for moderate values of $U$.
}
\label{Table:Exc}
\end{table}

\begin{figure}[bh]
%\includegraphics[width=1\linewidth]{lattice2.eps}
%\vskip -1.5cm
\includegraphics[width=1.0\linewidth]{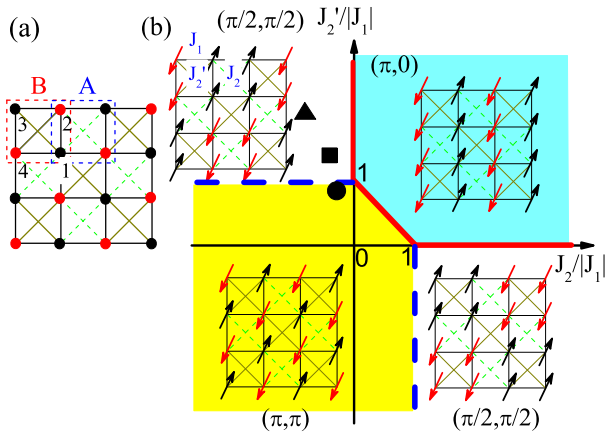}
\caption{(Color online) (a):
The doubled checkerboard lattice for the Hamiltonian defined
in Eq.~(\ref{Eq:HamHeisenberg}). It consists of two sets of
inter-penetrating plaquettes, indicated as blue (A) and red (B) dashed
squares. It can be decomposed into two (red and black circle)
sublattices.
%The underlying square lattice consists of two sublattices
%indicated as occupied black and red circles.
(b): The $T=0$ phase diagram.
% of this model. Three magnetically ordered states, ordered
%at $(\pi,\pi)$, $(\pi,0)/(0,\pi)$, and $(\pi/2,\pi/2)$,
%respectively, (see text) are found to be stabilized for $J_1>0$.
%Red solid lines indicate phase boundaries where the system is
%in absence of any long-range magnetic order; whereas blue dashed
% lines are boundaries showing first-order transitions.
The occupied symbols mark the ground states with exchange couplings listed in Table~\ref{Table:Exc}, with $U=1.5$ eV (triangle); $U=3.0$ eV (square); $U=4.5$ eV (circle).}
\label{FIG:phd}
\end{figure}

%In order to to better understand the DFT results,
%we 
We now 
turn to 
%the ground-state
%phase diagram of 
the effective
$J_1$-$J_2$-$J_2'$ Heisenberg model defined
on the two-dimensional doubled
checkerboard lattice shown in Fig.\ \ref{FIG:phd}(a).\
The classical
% $J_1$-$J_2$-$J_2'$ Heisenberg 
Hamiltonian is
$H=\sum_{\langle ij \rangle} J_{ij} \mathbf{S}_{i} \cdot \mathbf{S}_{j}$,
where $\mathbf{S}_{i}$ is a 3D vector with magnitude $S$.
%The 
To study the 
$T=0$ phase diagram,
% is obtained by finding the minimum of the 
%ground-state energy. For 
for $J_1>0$, 
%this is easily done by rewriting the summation in 
we rewrite the Hamiltonian 
via a summation over
two sets of inter-penetrating plaquettes shown in Fig.~\ref{FIG:phd}(a):
\begin{eqnarray}
\label{Eq:HamHeisenberg}
H &=& -(J_2+J'_2)NS^2 + \frac{1}{2}\sum_{m=A,B} \sum_{\square m} \{ J_{1m} (\sum_{i=1}^4\mathbf{S}_i)^2\nonumber\\
&+& J_{2m} [(\mathbf{S}_1+\mathbf{S}_3)^2+(\mathbf{S}_2+\mathbf{S}_4)^2]\},
\label{H-J-Plaquette}
\end{eqnarray}
where $J_{1A}=\alpha J_1$, $J_{1B}=(1-\alpha) J_1$; $J_{2A(B)}=J_2(J'_2)-J_{1A(B)}$.
%\qsx{Here $0\leqslant\alpha\leqslant1$, is a control parameter,}
Here $0\leqslant\alpha\leqslant1$ is a 
dummy control parameter~\cite{alpha-note},
and 
$N$ refers to the total number of plaquettes.
The phase diagram is shown in Fig.~\ref{FIG:phd}(b).
%As in the square lattice $J_1$-$J_2$ model, 
%the phase diagram shows competition between
%there 
There are
the N\'{e}el ordered AFM [for $0<\alpha<1$, $J_2<\alpha 
J_1, J'_2<(1-\alpha)J_1$] and the collinear
$(\pi,0)/(0,\pi)$ ordered AFM [for $0<\alpha<1$, 
$J_2>\alpha J_1, J'_2>(1-\alpha)J_1$] states.
%Besides these two states, a 
%In addition, 
%A third 
Unlike the square lattice $J_1$-$J_2$ model, 
there also exists
a plaquette $(\pi/2,\pi/2)$ 
AFM phase, which 
%ordered at $(\pi/2,\pi/2)$ 
is stable for $J_2(J'_2)>J_1$ and $J'_2(J_2)<0$. 
The 
%phase 
transition between any two
of the four ordered
% regimes 
regions in the phase diagram is first order.
% because of the different ordering vectors in these phases. 
Interestingly, we find that along the red solid boundaries
in Fig.~\ref{FIG:phd}(b),
% the ground state of 
the model loses any long-range magnetic order
since
%there 
the ground-state 
is infinitely degenerate. 
%This can be easily checked, for instance,
%for $\alpha=1$, $J'_2=0$, and $J_2>J_1$, 
%where the lattice consists of arrays of uncorrelated AFM
%spin chains along $J_2$ bonds.
Along this boundary, $J_2=J'_2=1/2$ corresponds to the maximally frustrated point of the $J_1$-$J_2$ model,
and $J_2(J'_2)=0$ 
%just restores 
corresponds to 
the checkerboard AFM model.
The phase diagram for $J_1<0$ can also be obtained by rotating all the spins in one sublattice of the underlying square lattice (see Fig.~\ref{FIG:phd}(a))
$\mathbf{S}_i\rightarrow-\mathbf{S}_i$ and sending $J_1\rightarrow-J_1$. It is similar to the one for $J_1>0$. The only
difference is that the N\'{e}el ordered $(\pi,\pi)$ phase is
% substituted
replaced by the FM
phase ordered at $(0,0)$.

%Knowing the phase diagram of the effective model, we discuss
%the DFT results at $U>0$ in Table~\ref{Table:Energy} and \ref{Table:Exc}.
The magnetic ground states expected from the exchange couplings 
given in Table~\ref{Table:Exc} are marked in Fig.~\ref{FIG:phd}(b).
For $U=1.5$ eV and $U=3.0$ eV, the ground state is the
$(\pi/2,\pi/2)$
plaquette
AFM state, while for larger $U$, i.e., $U=4.5$ eV, the
standard
$(\pi,\pi)$
N\'{e}el
state is energetically favorable. %%Considering the stacking
%%%structure of the material, these
Taking into account the stacking structure of the material,
we see that 
these results are consistent with the DFT energies listed in
Table~\ref{Table:Energy}: the AFM6
state has the lowest energy for $U=1.5$ eV
and $U=3.0$ eV, and the AFM1 state has the lowest
energy for $U=4.5$ eV.

To summarize, we studied the iron oxychalcogenides
La$_2$O$_2$Fe$_2$O(Se,S)$_2$,
whose Fe-square lattice is expanded compared to that
of the usually considered iron pnictides and chalcogenides.
We have theoretically demonstrated that their
Fe $3d$-electron bands are narrower than the usual
cases.
The corresponding enhancement of
correlation effects promotes the Mott insulating state,
and this is demonstrated experimentally.
%\qsx{Our results provide evidence that
%the parent systems of the iron pnictides and chalcogenides
%are already on the verge of Mott localization,
%and raise
%the prospect of systematic studies of the iron based systems
%across a Mott transition.}
Our 
%work is the first 
demonstration of a Mott insulating
phase in systems closely related to the iron pnictides
%, thereby extending 
extends the electronic behavior of these systems to a new regime.
%and raising the prospect of studying the iron-based materials 
%across a Mott transition. 
In addition, our results support the
notion that the iron pnictides/chalcogenides possess intermediately
strong electron correlations and are not too far away from 
Mott localization.

%\begin{acknowledgments}
%{\bf Acknowledgments.}
This work was supported by the National Nuclear Security Administration
of the U.S. DOE  at 
LANL 
under Contract No. DE-AC52-06NA25396, the U.S. DOE Office of Science,
and the LDRD Program at LANL (J.-X.Z.), 
the NSF Grant No. DMR-0706625, the Robert A. Welch Foundation
Grant No. C-1411, and the W. M. Keck Foundation
(R.Y. and Q.S.),
the NSFC Grant No.10974175 and 10874147, the National
Basic Research Program of China Grant No. 2009CB929104, and the PCSIRT of China
Contract No. IRT0754 (H.W., J.D. and M.F.),
and DoD MURI (L.L.Z. and E.M.). It was also supported in part by the Cyberinfrastructure for Computational Research funded by NSF under Grant CNS-0821727.
J.-X.Z. thanks L. Cario for correspondence and the IT team
at the Rice RCSG %Research Computing Support Group 
for
help with computational resources.

\end{document}